\documentclass[a4paper,aps,pra,showpacs,superscriptaddress,floatfix,amssymb,amsmath]{revtex4}
\usepackage{graphicx,dcolumn}
\begin{document}

\title{Search for anisotropic effects of hcp solid helium on optical lines of cesium impurities}

\author{M.~Melich}
\email{mathieu.melich@lkb.ens.fr}
\affiliation{Laboratoire Kastler Brossel, {\'E}cole Normale
Sup{\'e}rieure; CNRS; UPMC; 24 rue Lhomond, F75005 Paris, France}
\author{J.~Dupont-Roc}
\author{Ph.~Jacquier}
\affiliation{Laboratoire Kastler Brossel, {\'E}cole Normale
Sup{\'e}rieure; CNRS; UPMC; 24 rue Lhomond, F75005 Paris, France}

\begin{abstract}
The anisotropic effect of a hcp $^4$He solid matrix on cesium atoms has been proposed as a tool to
reveal the parity violating anapole moment of its nucleus. It should also result in splitting the
D$_2$ optical excitation line in a way depending on the light polarization. An experimental
investigation has been set up using oriented hcp helium crystals in which cesium metal grains are
embedded. Atoms are created by laser sputtering from this grains. Optical absorption spectra of the
D$_2$ line have been recorded in the temperature range of 1.0 to 1.4~K at liquid/solid coexistence
pressure by monitoring the fluorescence on the D$_2$ line at 950~nm. No significant effect of the
light polarization has been found, suggesting a statistically isotropic disordered solid
environment for the cesium atoms.
\end{abstract}


\maketitle
\section{Introduction}

Optical properties of heavy alkali metal atoms (Cs, Rb) embedded in liquid or solid helium have
been studied since 1990 \cite{1993} and reviewed in several articles \cite{revues}. Liquid helium and
b.c.c solid behave as isotropic matrices for the atomic impurities. On the contrary, uniaxial
h.c.p. solid is expected to perturb the atomic levels as a quadrupolar hamiltonian, lifting the
degeneracy between Zeeman sublevels with different values of $|m|$. Such a feature was indeed
observed experimentally in the two ground state hyperfine sublevels \cite{kanorsky1998} and was
interpreted as an anisotropic hyperfine interaction induced by the crystal. This effect was later
proposed as a tool to reveal the parity violating anapole moment of the cesium
nucleus \cite{bouchiat2001}. An electric and a magnetic field with well defined orientations with
respect to the helium crystal axis should produce a linear Stark effect, proportional to this
anapole moment. This experiment requires a uniform crystal orientation throughout the sample. This
remains to be achieved, since the experiments in the Cs ground state quoted above were interpreted
as resulting from a statistically isotropic distribution of the local crystal c-axis. Also unknown
is the perturbation of the local crystal order around each Cs atom. Optical absorption spectra are
a sensitive tool to test the isotropy of the local environment of the alkali atoms in the solid
matrix. We show that for an anisotropic matrix the shape of the D$_2$ absorption line is expected
to depend on the excitation light polarization, as well as that of the detected fluorescent light.
In addition we report the result of an experimental investigation on cesium atoms sputtered in an
initially mono-domain hcp crystal.

\section{Cesium impurity absorption spectrum in liquid and solid helium}

\begin{figure}[th]
\begin{center}
\includegraphics[scale=0.40]{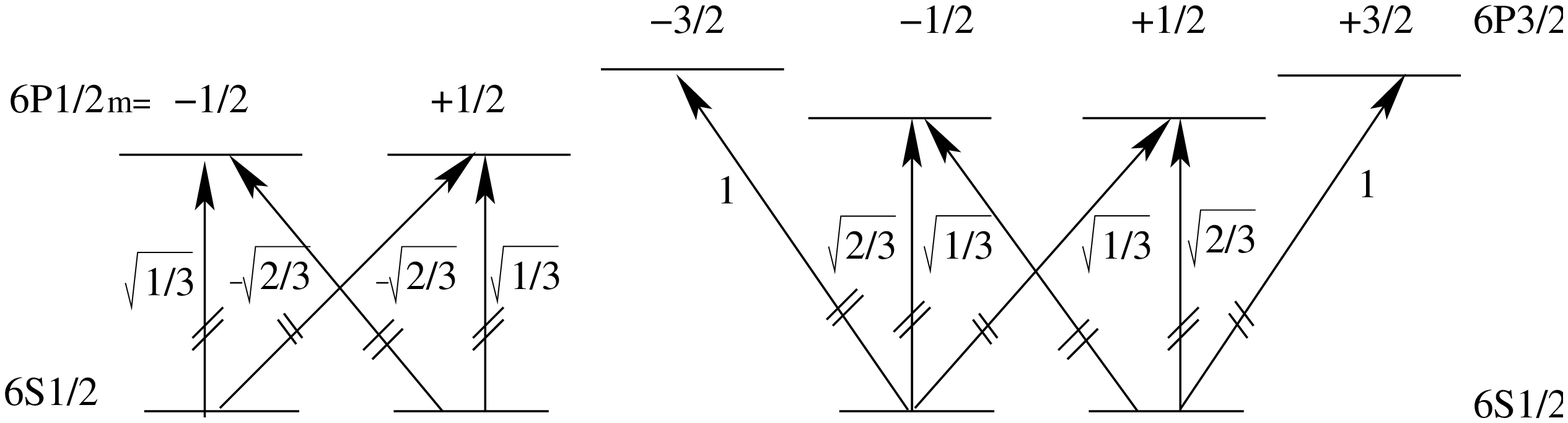}
\end{center}
\caption{ Energy levels of cesium with a quadrupolar perturbation. Hyperfine structure is ignored.
Numbers shown are proportional to the electric dipole matrix elements. The energy gaps between the
various energy levels are not to the same scale. Ground state is duplicated for
clarity.}\label{fig1}
\end{figure}

Optical transitions of a cesium atom embedded in a dense phase of helium have been described using
various models \cite{theory}. Neglecting the hyperfine structure, the electronic energy levels
reduce to the 8 sublevels shown in Fig.~\ref{fig1}. The D$_1$ line corresponding to the
6S$_{1/2}\rightarrow $6P$_{1/2}$ transition is simply broaden by the interaction with the
surrounding helium atoms. The properties of the D$_2$ line (6S$_{1/2}\rightarrow$ 6P$_{3/2}$
transition) is due to the removal of the 4-fold degeneracy of the 6P$_{3/2}$ level by any
anisotropic environment, resulting into two Kramers doublets.
Let us first ignore fluctuations and assume that the environment has an axial symmetry, as the hcp
crystal does. Then taking the $z$-axis along this direction, its action on the  6P$_{3/2}$ level
 may be written as ${\cal H} = A + B \left(3 J_z^2 - J(J+1)\right)/6$. The energy sublevels are
 the $m=\pm 3/2$ and $m=\pm 1/2$ magnetic sublevels with respect to the
z-axis, with energy $A+B/2$ and $A-B/2$ respectively. The matrix elements of the electric dipole
moment responsible for the optical transitions are reminded in Fig.~\ref{fig1}. They imply that the
weights of the two absorption line components strongly depend on the light polarization with
respect to the $z$-axis. For z-polarized light, only the $|{1/2}|$ component is excited, while for
$x$- or $y$-polarization, the ${3/2}$ and ${1/2}$ components are excited with the probability 1/2
and 1/6 respectively. Experimentally, the absorption of light is very weak.  Since radiative decay
is the dominant decay process for the 6P excited state, one measures instead the intensity in the
atomic fluorescence band as a function of the excitation wavelength. This intensity depends also on
fluorescence polarization.  Table~I gives relative line intensities for various polarizations of
fluorescence light. It is implicitly assumed that relaxation in the excited state prior to
fluorescence preserve the populations of the $m$ sublevels. This amounts to neglect  effects of the
hyperfine coupling  in the excited state. A fourth case is also considered which assumes on the
contrary a complete depolarization of the excited state before fluorescence. It appears clearly
that in any case the absorption spectrum strongly depends on excitation and detection
polarizations.
\begin{table}[h]
\begin{center}
\begin{tabular}{|c|cc|cc|}
\hline %
Excitation & \multicolumn{2}{c|}{z-polarization} & \multicolumn{2}{c|}{y-polarization}
\\
\hline\hline %
Fluorescence detection & $I_{3/2}$ &  $I_{1/2}$ & $I_{3/2}$ & $I_{1/2} $ \\ %
 z-analyzer & 0 & 4/9 & 0 & 1/9   \\ %
 x-analyzer & 0 & 1/9 & 1/4 & 1/36   \\ %
 no analyzer & 0 & 5/9 & 1/4 & 5/36   \\ %
 depolarized  & 0 & 4/9 & 1/3 & 1/9  \\ %
\hline
\end{tabular}
\caption{Intensities of the two components $I_{3/2}$ and $I_{1/2}$ of the D$_2$ absorption line
excited by a beam on the x-axis and detected on the fluorescence intensity in the y-direction for
various polarizations.}\label{tab1}
\end{center}
\end{table}
Consider now the effect of dynamic fluctuations. The effect of such a perturbation has been
considered by Kinoshita {\it et al.} in the case of liquid helium \cite{kinoshita1996}, with no
static anisotropy. Because of the three dimensions of space, an isotropic perturbation is less
probable than anisotropic ones, and the line is split into two broad components of equal weights.
Because of the statistical isotropy, no effect of polarization on the two component weights is
expected. The more general case with a remaining static anisotropy is more complicated. It can be
shown to {\it reduce} the contrast between the two components. If the static anisotropy is not
small with respect to the fluctuations, some effect of light polarization should remain. This
motivated the following experimental investigation.

\section{Experimental arrangement and procedure}

The experimental cell is a 4~cm stainless steel cube, with five silica windows (diam. 2.5~cm),
cooled from the top by a pumped helium-4 fridge in the 1.0-1.4~K temperature range. To achieve
single crystal growth, the nucleation and growth is made at constant temperature and pressure
(1.2~K and 25.5~bar). An electro-crystallization device \cite{keshishev1979} provides a unique seed
which falls on the bottom of the cell. The crystal is subsequently grown from this seed. The
orientation of the {\it c-}facet (most often horizontal) is easily monitored visually during the
growth process which takes place below the corresponding roughning transition
1.3~K \cite{balibar2005}. Cesium atoms are produced in two steps by laser sputtering from a metal
target \cite{1993} located in the upper part of the cell filled with superfluid. The laser pulses
(1-5~mJ, 8~ns, $\lambda$=532~nm) are provided by a Nd-YAG laser. The first step produces metallic
grains, about 100~$\mu$m in size, which fall on the crystal surface, when the interface is located
in the middle of the windows. These grains are then incorporated into solid helium by subsequent
growth. In the second step, the laser is focused on these grains, and atoms are produced in their
immediate vicinity. This production phase lasts between a few minutes to half an hour at a rate of
2 pulses/second with the aim of maximizing the fluorescence signal on the D$_1$ line. The D$_1$ and
D$_2$ lines are excited by a tunable Ti:Sa CW laser scanned between 770 and 870~nm, with a $z$- or
$y$- polarization, the z-axis being along the vertical c-axis of the grown crystal. The
fluorescence light  in the 950~nm band \cite{muller-sibertPhD,nuttels2005} is detected by a cooled
APD and selected along the $y$-axis by a 900~nm-edge filter which blocks the stray light from the
laser. The other 880~nm band is also occasionally used to monitor the D$_1$ fluorescence.

\section{Results}

A first observation is the time $T_a$ over which cesium atoms can be observed after their initial
preparation. A very significant extension of $T_a$ was obtained when lowering temperature from
1.4~K to 1.0~K. While at 1.4~K, fluorescence from Cs atoms disappears in a few hours, as observed
in Weis' group \cite{arndt1995}, at low temperatures it can be kept for days and even weeks. Because
the diffusion of impurities in solid helium is expected to be thermally activated in this
temperature range, such a behaviour is not unexpected. It is however of primary importance for
future experiments with Cs atoms which will need long integration times.

\begin{figure}[thb]
\begin{center}
\includegraphics[scale=0.55]{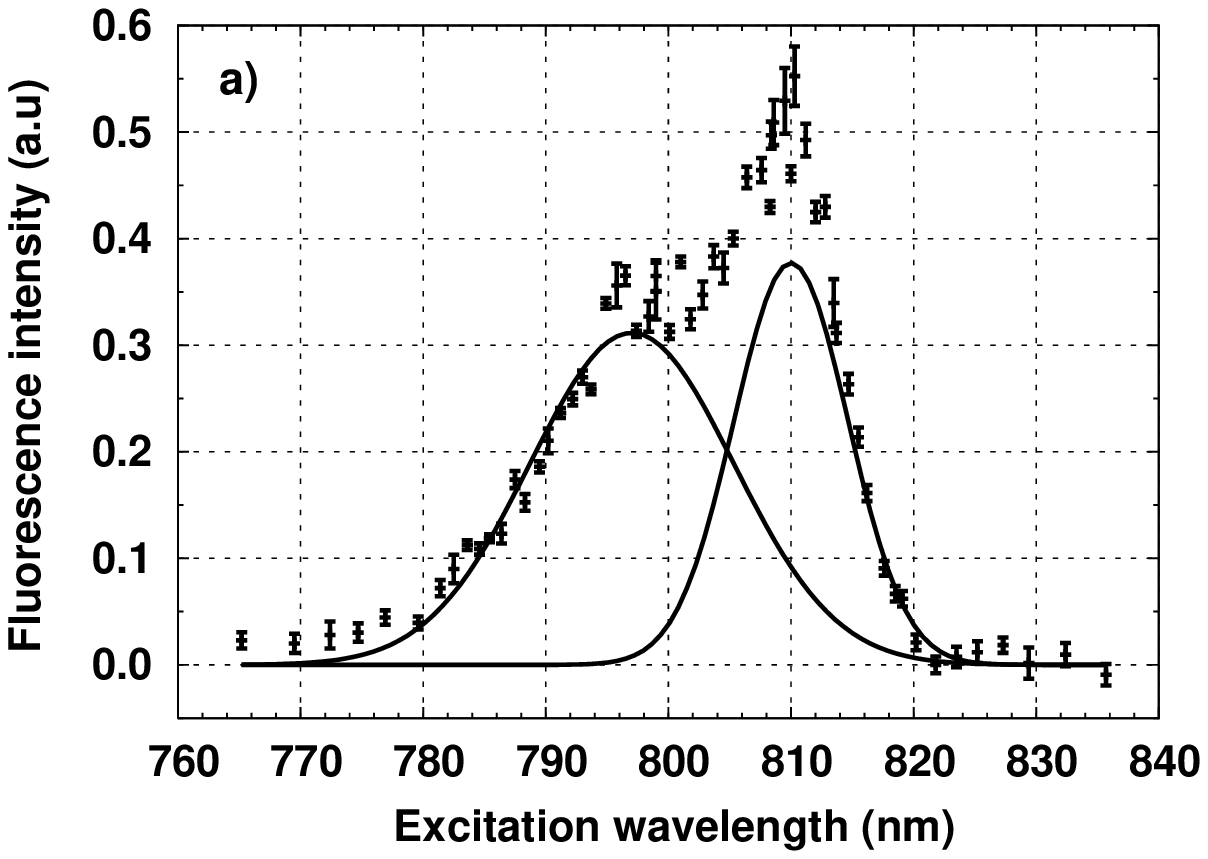}
\includegraphics[scale=0.55]{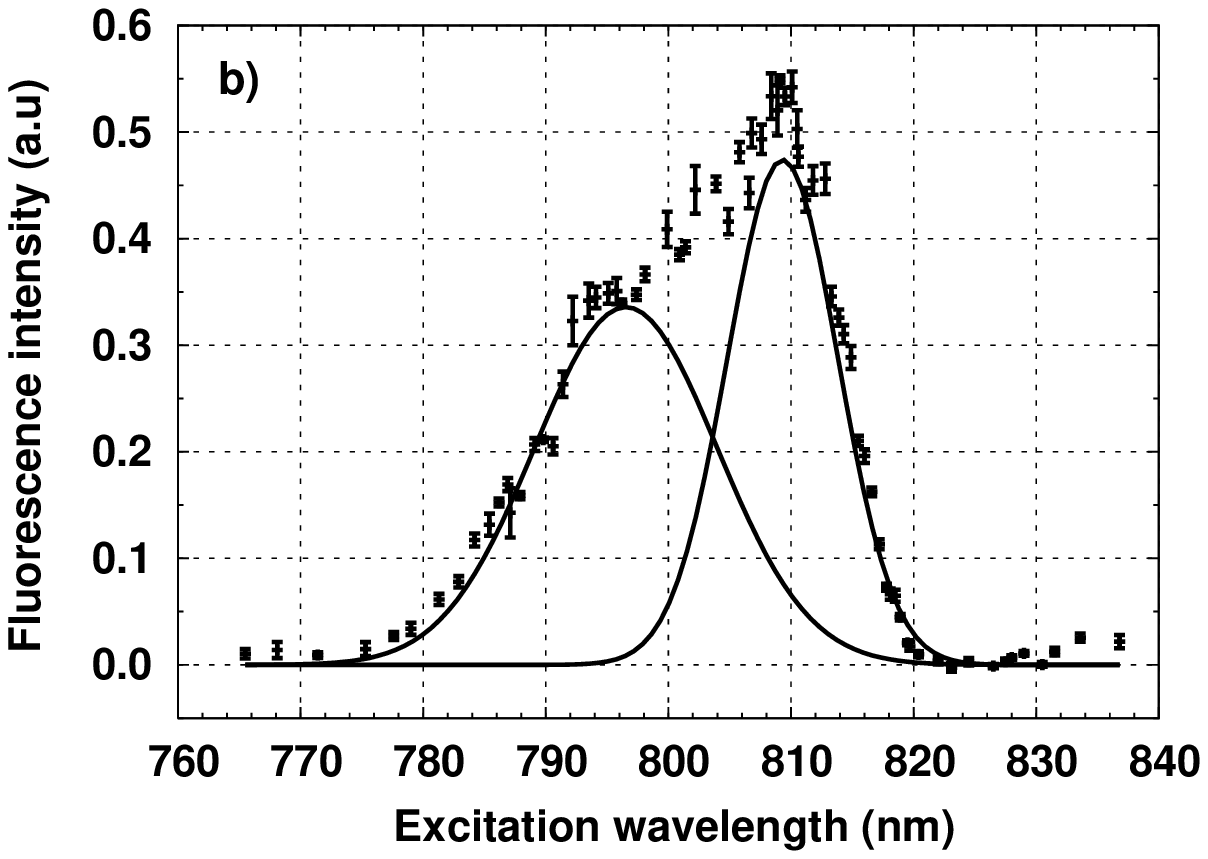}
\end{center}
\caption{Absorption spectra of the D$_2$ line for different polarizations. (a)~$z$-excitation,
$z$-analyzer on detection, (b)~$y$-excitation, $x$-analyzer on detection. Fitting the spectra with
a sum of two gaussian lines yields the ones shown as plain curves.}\label{fig3ab}
\end{figure}

Absorption spectra of the D$_2$ line are shown in Fig.~\ref{fig3ab} for various polarizations of
excitation and detection. A splitting of the line into two components is clearly visible. Its
magnitude corresponds to an energy gap of 200$\pm 10$ cm$^{-1}$.  The line shape is similar to what
has been reported for liquid helium \cite{kinoshita1996} and solid helium at higher
pressure \cite{moroshkin2006b}. Quantitatively, line widths are about 15\% larger than those in the
liquid at 20~bar while splittings are 20\% larger.
 The effect of the excitation polarization on the line shape is much
smaller than expected. More precisely, the component weight ratio is 0.7 for $z$-excitation,
$z$-analyzer and 0.85 for $y$-excitation, $x$-analyzer, to be compared with 0 and 9 expected
respectively (see Table~\ref{tab1}).

\section{Discussion}

There are two ways for explaining this result. The first one is to conclude that there is no
permanent anisotropy in the environment of the atoms and that the splitting observed is mostly due
to dynamic anisotropic fluctuations similar to those observed in liquid helium \cite{kinoshita1996}
as already discussed in reference \cite{moroshkin2006b}. The similarity of line shapes is indeed
striking. Whether such an explanation is compatible with the hyperfine and magnetic resonance
spectra observed in the ground state remains to be investigated. The second explanation is the one
put forward in reference \cite{kanorsky1998}~: the sputtering process destroys the global order of
the crystal which is transformed in a mixture of randomly oriented small hcp domains. Then no
effect of the light polarization remains. If so, improvement of the crystal quality is to be
searched for. Partial fusion, and re-crystallization from an ordered part of the solid is to be
tried. It is unclear whether atoms will survive to such a process. Less damaging sputtering process
using a femtosecond laser could also be attempted.\\


\begin{acknowledgments}
We acknowledge support from ANR, grant 05-BLAN-0084-01.
\end{acknowledgments}



%



\end{document}